\documentclass[preprint,floatfix
,aps,prd,superscriptaddress]{revtex4}

\usepackage[utf8]{inputenc}
\usepackage{graphicx}
\usepackage{tikz}
\usepackage{amssymb,amsmath,mathtools,comment,hyperref}
\usepackage{mathrsfs}
\usepackage{xcolor}

\usepackage{tensor}

\newcommand{\ket}[1]{|#1\rangle}
\newcommand{\bra}[1]{\langle #1|}

\newcommand{\mo}{\mathcal{O}}
\newcommand{\T}{\theta}
\newcommand{\D}{\Delta}
\newcommand{\s}{\sin}
\newcommand{\C}{\cos}

\begin{document}

\title{
Acceleration radiation of an atom freely falling 
into 
\\
a Kerr black hole and near-horizon conformal quantum mechanics}
\author{A. Azizi}
\affiliation{Institute for Quantum Science and Engineering, Texas A\&M University, College Station, Texas, 77843, USA}
\author{H. E. Camblong}
\affiliation{Department of Physics and Astronomy, University of San Francisco, San Francisco, California 94117-1080, USA}
\author{A. Chakraborty}
\affiliation{Department of Physics, University of Houston, Houston, Texas 77024-5005, USA}
\author{C. R. Ord\'{o}\~{n}ez}
\affiliation{Department of Physics, University of Houston, Houston, Texas 77024-5005, USA}
\affiliation{Department of Physics and Astronomy, Rice University, MS 61, 6100 Main Street, Houston, Texas 77005, USA.}
\author{M. O. Scully}
\affiliation{Institute for Quantum Science and Engineering, Texas A\&M University, College Station, Texas, 77843, USA}
\affiliation{Baylor University, Waco, TX 76706, USA}
\affiliation{Princeton University, Princeton, New Jersey 08544, USA}

\date{\today}
\begin{abstract} 
 An atom falling freely into a Kerr black hole in a Boulware-like vacuum 
 is shown to emit radiation with a Planck spectrum at the Hawking temperature. For a cloud of falling atoms with random initial times, the radiation is thermal. The existence 
 of this radiation is due to the acceleration of the vacuum field modes with respect to the falling atom.
Its properties can be traced to the dominant role of 
 conformal quantum mechanics (CQM) in the neighborhood of the event horizon. 
We display this effect for a scalar field, though the acceleration radiation
has a universal conformal behavior that is exhibited by all fields in the background of generic black holes. 

\end{abstract}

\maketitle

\section{Introduction}
\label{sec:intro}
\noindent
Hawking's seminal work on black hole radiance through quantum processes~\cite{hawking1974,hawking1975} has given rise to a research area that also includes black hole thermodynamics~\cite{Bekenstein_entropy,hawking_BHmech}. A related phenomenon was discovered by Unruh \cite{unruh76}, following earlier work by Fulling and Davies~\cite{fulling73,davies74}, and showing that an accelerated observer in flat spacetime detects particles in the Minkowski vacuum. Additional insights into the Hawking and Unruh effects, and black hole thermodynamics, are of great interest due to their apparent universality, and as a litmus test of any candidate theories of quantum gravity.
\\ \indent
  In this paper, we probe an aspect of the deep connections between Hawking and Unruh radiation via a gedanken experiment where an atom falls freely into a Kerr black hole in a Boulware-like vacuum and emits radiation with a spectrum similar to black hole radiance. We generalize
the insightful quantum-optics setup of Refs.~\cite{scully2018, scully2019} 
(and related work~\cite{Chakraborty-Majhi_2019,Dalui-Majhi-Mishra_2020}),
as well as the conformal derivation of Ref.~\cite{camblong2020},
from the Schwarzschild to the Kerr metric.
The Kerr geometry is the solution of the 4D vacuum Einstein field equations with a rotating black hole of mass $M$ and angular momentum $J$, given by
\begin{align}
ds^2 = -\frac{\D}{\rho^2}(dt-a\s^2 \T d\phi)^2 
+ \frac{\rho^2}{\Delta}dr^2 + \rho^2d\T^2 
+ \frac{\s^2\T}{\rho^2}\left[(r^2+a^2)d\phi-a dt\right]^2 
\label{eq:Kerr2}
\; 
\end{align}
in Boyer-Lindquist coordinates $(t,r,\theta,\phi)$.
In Eq.~(\ref{eq:Kerr2}), $a=J/M$ is the Kerr parameter, the auxiliary variables 
\begin{equation}
\D = r^2 -2Mr + a^2 \; \; \;  \text{and} \; \; \;  \rho^2 = r^2 + a^2 \C^2\T
\end{equation}
are introduced, and we use geometrized units $c=1, G= 1$.
This extension is of crucial relevance because such black holes: (i) provide models that closely match astronomical observations of gravitationally collapsed objects with angular momentum that generate gravitational waves~\cite{LIGO:2016}; (ii) are conceptual laboratories that test the robustness of extreme-gravity effects that are not artifacts of the spherical symmetry of the Schwarzschild solution. 
  \\ \indent
  With this background in mind, we will show that these radiation processes are driven by conformal symmetry, in which a part of the system looks identical under arbitrary magnifications near the outer event horizon~\cite{camblong2020}. These findings display the connections between black hole thermodynamics and horizon conformal symmetries that provide a statistical foundation for the Bekenstein-Hawking entropy~\cite{carlip1,carlip2,strominger}. The relation of a 2D conformal field theory (CFT) with the near-horizon asymptotic symmetries has been studied for extremal and near-extremal black holes within the Kerr/CFT correspondence~\cite{stromingerkerr}--\cite{strominger10}. In addition, in Ref.~\cite{parkerprl}, the two-point functions of 2D CFT are used to derive the thermal radiance of a non-extremal Kerr black hole. Another aspect of near-horizon conformal behavior is conformal quantum mechanics (CQM)~\cite{Padmanabhan_BH-ISP,guptasen,nhthermocqm,nhcamblong-sc,vaidya,moretti}, which is also analyzed as a (0+1)-dimensional CFT~\cite{dff,HEC-CRO_CQM}; here, near-horizon CQM governs the thermodynamics of a Schwarzschild black hole through a singular statistical mode counting that requires renormalization~\cite{nhthermocqm,nhcamblong-sc}. Thus, we will show that: (i) the non-extremal Kerr geometry exhibits an asymptotically exact near-horizon CQM that extends the scope of Refs.~\cite{nhthermocqm,nhcamblong-sc,tightnesscamblong} and (ii) acceleration radiation is created by free fall into a non-extremal Kerr black hole in a Boulware-like vacuum, with a dominant CQM contribution  
 that resembles a thermal spectrum at the Hawking temperature.
   This confirms that robustness of the results of Refs.~\cite{scully2018,scully2019,camblong2020} and \cite{nhthermocqm,nhcamblong-sc,tightnesscamblong}---beyond spherical symmetry---and highlights the universality of near-horizon CQM.
   
This remainder of this paper is organized as follows. 
In Sec.~\ref{sec:K-G_Kerr-nh}, 
we study the scalar field equations in general, and in their near-horizon CQM form.
In Sec.~\ref{sec:excitation-free-fall_Kerr},
we consider the interaction between the field and an atom, and the ensuing probabilities.
The near-horizon spacetime trajectories are analyzed in Sec.~\ref{sec:nh-spacetime-trajectory}.
Section~\ref{sec:nh-excitation-probability}
deals with the final expressions for the emission and absorption probabilities and the thermal radiation properties.
The article ends with the conclusions in Sec.~\ref{sec:discussion},
followed by the appendices: on the derivation of the near-horizon CQM equations
(\ref{app:CQM_derivation})
and the vacuum states (\ref{sec:field-modes-vacuum}).

\section{Klein-Gordon equation in Kerr geometry and near-horizon CQM}
\label{sec:K-G_Kerr-nh}
\noindent
In the Kerr geometry of Eq.~(\ref{eq:Kerr2}), 
we will consider the non-extremal case, with $M>a$, for which 
 $\D'_{+} \equiv \D'(r_{+}) = r_+-r_- \neq 0$
 (and where the prime denotes radial derivative). 
The outer ($r_+$) and inner ($r_-$) horizons of the black hole are given by the roots of the equation 
$\D =0$, i.e., $r_{\pm}= M\pm (M^2-a^2)^{1/2}$. 
The Kerr geometry is stationary and axisymmetric as the metric components are independent of the coordinates $t$ and $\phi$ respectively, with associated Killing vectors $\boldsymbol\xi_{(t)} = \partial_t$ and $ \boldsymbol\xi_{(\phi)} = \partial_\phi$. Equation~(\ref{eq:Kerr2}) can be rewritten in a more illuminating form
\begin{align}
 \! \!  ds^2 = -\frac{\D \rho^2}{\Sigma^2} dt^2 
 + \frac{\rho^2}{\Delta}dr^2 + \rho^2d\T^2    
+ \frac{\Sigma^2}{\rho^2} \s^2 \T \left( d \phi - \varpi dt \right)^2
\label{eq:Kerr3} 
\end{align}
with 
\begin{equation}
\Sigma^2 = (r^2 +a^2)^2 - \D \, a^2 \s^2 \T
 \; \; \;  \text{and} \; \; \;  
\varpi = - g_{t\phi}/g_{\phi \phi} 
\, .
\end{equation}
Equation~(\ref{eq:Kerr3}) describes a frame-dragging rotation with position-dependent angular velocity $\varpi$ relative to the external reference system~\cite{Landau-Lifshitz_CTFs}. This picture is mandatory within the ergosphere, which is limited by the largest root of $g_{tt}=0$, where
$ \boldsymbol\xi_{(t)} $ becomes spacelike. As the outer event horizon is approached, $\varpi$ becomes the angular velocity of the black hole,
\begin{equation}
\Omega_{H} =
\lim_{r \rightarrow r_{+}} \varpi
=
 \frac{a}{2M r_{+} }
 =
 \frac{a}{r_{+}^{2}+a^{2} }
\; .
\label{eq:n-h_expansion}
\end{equation}
\\ \indent
We will now consider the interaction of a real scalar field of mass $\mu_{\Phi}$ with an atom of mass $\mu$ in the gravitational background of a Kerr black hole. The scalar field satisfies the Klein-Gordon equation
\begin{align}
\left[ \Box - \mu_{\Phi}^{2} \right] \Phi \equiv \frac{1}{ \sqrt{-g} }\partial_{\mu} \left(
\sqrt{-g} \, g^{\mu \nu}\, \partial_{\nu} \Phi \right) - \mu_{\Phi}^{2} \Phi = 0\; .
\label{eq:Klein_Gordon_basic}
\end{align}
In addition, for a metric of the form~(\ref{eq:Kerr2}), Eq.~(\ref{eq:Klein_Gordon_basic}) reduces to 
\begin{equation}
-\frac{\Sigma^2}{\D} \frac{\partial^2 \Phi}{\partial t^2}
- \frac{4 M r a}{\D}  \frac{\partial^2 \Phi}{\partial t \, \partial \phi}
+ \left(\frac{1}{\s^2 \T} - \frac{a^2}{\D} \right) \frac{\partial^2 \Phi}{\partial \phi^2}
+ \frac{\partial}{\partial r} \left( \D  \frac{\partial \Phi}{\partial r} \right)
+ \frac{1}{\s \T}  \frac{\partial}{\partial \T} \left( \s \T  \frac{\partial \Phi}{\partial \T} \right) 
- \mu_\Phi^{2}  \rho^2 \Phi = 0 \; .
\label{eq:Kerr_Klein_Gordon_spatial}
\end{equation}
This scalar-field equation is a particular case (for spin $s=0$) of the Teukolsky equation~\cite{frolov}. Due to the
structural form of Eq.~(\ref{eq:Kerr_Klein_Gordon_spatial}), the resulting conformal behavior displayed below is universal, namely, it is exhibited by all fields (with arbitrary spin) in the background of generic black holes. 
\\ \indent
Equation~(\ref{eq:Kerr_Klein_Gordon_spatial}) can be studied via the separation of variables
\begin{equation}
\phi_{\omega l m} (\boldsymbol{r}, t) 
= R(r) S(\theta) \, e^{im{\phi}}  e^{-i{\omega} {t}}
\label{eq:KG_separation}
\end{equation}
(replacing $\Phi = \phi_{\omega l m} $ to specify the mode functions; see next section), 
where $\boldsymbol{r}=(r,\theta, \phi)$. Specifically, the functions
$R(r) \equiv R_{\omega l m} (r)$ and
$S(\theta) \equiv S_{\omega l m} (\theta)$ 
satisfy the equations given in Appendix~\ref{app:CQM_derivation}.
These functions depend on the frequency $\omega$ and the 
 quantum numbers $l$, $m$.
In particular, $S_{\omega l m} (\theta)$ 
 are oblate spheroidal wave functions of the first kind~\cite{spheroidal}.
Moreover,
 the coordinate change~\cite{parkerprl},
\begin{equation}
\Tilde{t}=t  
\; , \; \; \; 
\Tilde{\phi}=\phi-\Omega_H t
\; ,
\label{eq:rotating-coords_separation_1}
\end{equation}
with
\begin{equation}
\phi_{\tilde{\omega} l m} (\boldsymbol{r}, t) 
= R(r) S(\theta) e^{im\tilde{\phi}}  e^{-i\tilde{\omega} \tilde{t}}
\; , \; \; \; 
\tilde{\omega}=\omega-m\Omega_H
\; ,
\label{eq:rotating-coords_separation_2}
\end{equation}
defines a frame that is corotating with the black hole at the angular velocity $\Omega_{H}$,
and with shifted frequency $\tilde{\omega}$. The associated Killing vector  
$ \boldsymbol\xi_{(\tilde{t})} =  \boldsymbol\xi_{(t)} + \Omega_{H} \boldsymbol\xi_{(\phi)}$ is timelike
when sufficiently close to the event horizon.
The horizon $\mathcal{H}$ is a null hypersurface with respect to
this Killing vector
$\boldsymbol\xi_{(\tilde{t})} $; and
the associated surface gravity $\kappa = -(\nabla_{\mu} \xi_{\nu})( \nabla^{\mu} \xi^{\nu})/2$ (with $r=r_+$) 
takes the value 
\begin{equation}
\kappa= \frac{ \D'_{+} }{ 2(r_+^2+a^2) }
\; .
\label{eq:surface-gravity_Kerr}
\end{equation}
\\ \indent
We will systematically enforce the near-horizon expansion, 
denoted by $\stackrel{(\mathcal H)}{\sim}$,
 in terms of $x \equiv r-r_+ \ll r_+$. This involves the replacements
\begin{eqnarray}
\D(r)  \stackrel{(\mathcal H)}{\sim} & \D'_{+}  \, x \left[ 1 + O(x) \right]
\; , \; \; \; 
\D'(r)  \stackrel{(\mathcal H)}{\sim} & \D'_{+} \left[ 1 + O(x) \right]
\; , \; \; \; 
\D''(r)  =   \D'_{+}  =2 
\, .
\end{eqnarray}
 Then, the radial near-horizon leading order of Eq.~(\ref{eq:Kerr_Klein_Gordon_spatial}) is
\begin{equation}
\left[\frac{1}{x} \frac{d}{d  x} \left( x  \frac{d}{d x} \right) 
+ \left( \frac{ \tilde{\omega}}{2 \kappa} \right)^{2}
\frac{1}{x^2} \right] R(x)
\stackrel{(\mathcal H)}{\sim} 0 \; ,
\label{eq:Kerr_Klein_Gordon_conformal-R}
\end{equation}
as derived in Appendix~\ref{app:CQM_derivation}.
The correspondence with the standard form of CQM can be established with the Liouville transformation $R(x) \propto x^{-1/2} u(x)$, whence the near-horizon reduced radial function $u(x)$ satisfies
\begin{equation}
\frac{d^2 u(x) }{d  x^2} + \frac{ \lambda }{x^{2}} \,
\left[ 1 + \mathcal{O}(x) \right] u (x) = 0\;  ,
\label{eq:Kerr_Klein_Gordon_conformal}
\end{equation}  
where 
\begin{equation}
\lambda = \frac{1}{4} + \Theta^2\, , \hskip 4em \Theta = \frac{\tilde{\omega}}{2\kappa} 
\; .
\end{equation}
 The scale invariance of Eq.~(\ref{eq:Kerr_Klein_Gordon_conformal}) under a rescaling of $x$ can be seen from the form of the effective potential $V_{\rm eff} (x)= -\lambda/x^{2}$, such that the 1D Schr\"{o}dinger Hamiltonian $\mathscr{H} = p_x^2/2m + V_{\rm eff}(x)$ describes the strong coupling regime of CQM. This operator, along with the dilation operator $D$ and special conformal operator $K$ (defined using a time parameter conjugate to $\mathscr{H}$), produces the noncompact SO(2,1) algebra. A manifestation of this conformal symmetry is the disappearance of all characteristic field scales; in particular, $\mu_{\Phi}$  plays no role in the near-horizon physics. This algebraic structure, along with the singular near-horizon effective potential, have been shown to determine the thermodynamics of a Schwarzschild black hole~\cite{nhthermocqm}. The emergence of the CQM Hamiltonian for  the Kerr geometry shows the {\it universality\/} of near-horizon CQM. 
\\ \indent
A pair of linearly independent solutions to Eq.~(\ref{eq:Kerr_Klein_Gordon_conformal}) is given by $u(x)\propto  x^{1/2  \pm i\Theta}$. When combined with their time dependence from Eq.~(\ref{eq:rotating-coords_separation_2}), this yields
\begin{equation}
\Phi^{ \pm {\rm \scriptscriptstyle (CQM)} }_{\omega l m} \propto 
R^{ \pm {\rm \scriptscriptstyle (CQM)} }
 \, e^{im\tilde{\phi}}  e^{-i \tilde{\omega} \tilde{t} } 
\propto x^{\pm i \Theta} 
 e^{im\tilde{\phi}}  e^{-i \tilde{\omega} \tilde{t} } 
\label{eq:CQM_modes}
\; ,
\end{equation}
 which are outgoing/ingoing CQM modes
 normalized as asymptotically exact WKB local waves~\cite{nhcamblong-sc}. 
The $\tilde{\phi}$ dependence is kept for consistency in subsequent calculations of excitation probabilities in the presence of frame dragging in the near-horizon region. Incidentally, to leading order in the near-horizon expansion, $R_{\pm} (x) \stackrel{(\mathcal H)}{\sim}  e^{-i\tilde{\omega}(t\mp r_*)}$, where the tortoise coordinate $r_*$ for the Kerr metric is defined through the equation 
${dr_*}= f^{-1} {dr} $, where $f= {\D}/ {(r^2+a^2)} $.
This equivalence is shown in Appendix~\ref{app:CQM_derivation}.
In practice, our preferred use of the CQM modes makes the near-horizon conformal behavior more explicit. In what follows, we will consider the outgoing modes to find the excitation probability for an atom falling freely towards a Kerr black hole in a Boulware-like vacuum state; as shown in Sec.~\ref{sec:nh-excitation-probability}, the ingoing modes do not contribute to the probability amplitude.
  \\ \indent
  The generic Boulware-like vacuum can be defined with respect to the Boyer-Lindquist coordinates in the rotating frame~(\ref{eq:rotating-coords_separation_1}). 
Such a choice is similar to the Boulware vacuum associated with ordinary Schwarzschild coordinates in the Schwarzschild geometry; however, as the Kerr geometry is stationary, with significant frame dragging, the relevant coordinates~(\ref{eq:rotating-coords_separation_1}) are adapted to the black hole rotation.
As discussed in
Appendix~\ref{sec:field-modes-vacuum}, the standard inclusion of the asymptotic regions, past/future null infinity $\mathscr{I}^{\mp}$, implies the existence of superradiant modes that complicate the ordinary definition of the vacuum---possible constructions that circumvent this issue are discussed therein.
Alternatively, the asymptotic regions at $\mathscr{I}^{\pm}$ can be replaced with bounded-domain boundary conditions (e.g., Dirichlet)  by enclosing the system within a boundary or mirror $\mathcal{M}$ placed inside the speed-of-light (SOL) surface~\cite{rotating-Qvacuum}.
The SOL surface is defined as that outside which an observer can no longer have the angular velocity $\Omega_{H}$ and $\boldsymbol\xi_{(\tilde{t})} $ becomes null.
 Then, the superradiant modes and the Unruh-Starobinsky effect would be removed, yielding a unique Boulware vacuum $\left| B_{\mathcal M} \right\rangle$,
 as well as a Hartle-Hawking vacuum analog~\cite{mirror-no-superradiance}---these are the natural states adapted to a frame corotating with angular velocity $\Omega_{H}$. 

\section{Excitation of a freely falling atom into the Kerr black hole}
\label{sec:excitation-free-fall_Kerr}
\noindent
We consider a two-state atom falling freely into the black hole. 
This problem can be tackled with an approach similar to that of Refs.~\cite{scully2018,scully2019,camblong2020}, with the field in a Boulware-like vacuum.
(See Appendix~\ref{sec:field-modes-vacuum}
for a discussion of modes and vacuum states.)
 The atom's Hamiltonian is given by $\mathscr{H}_{\rm at} = \left(\ket{a}\bra{a} - \ket{b}\bra{b}\right) \nu/2\,$  (with $\hbar=1$), where $\ket{a}$ and $\ket{b}$ are the atom's excited and ground states respectively, and $\nu$ is the atomic transition frequency. 
 The field operator is expanded with a complete set of orthonormal modes  $ \phi_{\boldsymbol{s}} (\boldsymbol{r}, t) $ as 
 \begin{equation}
      \Phi = \sum_{\boldsymbol{s}}
      \left[ a_{\boldsymbol{s}} 
     \phi_{\boldsymbol{s}} (\boldsymbol{r}, t) 
           + 
    \mathrm{h.c.} \right] 
    \; ,
    \label{eq:field-operator}
\end{equation}
where
 $a_{\boldsymbol{s}}$ is the field lowering operator that annihilates the vacuum,
and the field quanta are scalar (spin-0) ``photons.'' 
The symbol $\mathrm{h.c.}$ stands for the Hermitian (adjoint) conjugate.
 The atom-field interaction is treated as a weak dipole coupling in the interaction picture, with
\begin{equation}
    \hat{V}_I(\tau) = g\left[ {a}_{\boldsymbol{s}}
      \phi_{\boldsymbol{s}} (\boldsymbol{r}(\tau), t (\tau) )
     + h.c. \right] \!
    \left( \sigma_{\!_-} e^{-i \nu \tau} + \mathrm{h.c.} \right) 
    \; ,
    \label{eq:interaction-potential}
\end{equation}
where 
 $\sigma_{\!_-}$ is the atom's lowering operator, 
$g$ is the interaction strength, and $\tau$ is the atom's proper time.
In Eqs.~(\ref{eq:field-operator}) and
(\ref{eq:interaction-potential}), the symbol ${\boldsymbol{s}}$ stands for the set of ``quantum numbers'' that provide complete characterization of the mode: 
it includes the frequency $\omega$ of the mode and any additional numbers associated with the geometry and separation of variables.
For the Kerr geometry in 3 spatial dimensions, this is
${\boldsymbol{s}} =  \{\omega,l,m\}$, where 
$\{l,m\}$ are the spheroidal number and the ``magnetic'' quantum number associated with angular momentum. 
(When quantization is enforced in a finite box, the frequencies involve a third discrete number. Also,
for the particular case of the Schwarzschild geometry, these reduce to the usual numbers associated with spherical symmetry and angular momentum.)

Equation~(\ref{eq:interaction-potential})
 allows for free-fall virtual processes of the atom transitioning 
from the ground state $\ket{b}$ to the excited state $\ket{a}$, 
and creating a field-mode quantum (state ${\boldsymbol{s}}$),
with an excitation probability~\cite{QED-accel-Scully-etal_1,QED-accel-Scully-etal_2}, which is the emission probability for the field mode ${\boldsymbol{s}}$, given by
\begin{equation}
P_{e, {\boldsymbol{s}} }  
    = 
    \left|\int d\tau\; \bra{1_{\boldsymbol{s}},a} \hat{V}_I(\tau) \ket{0_{\boldsymbol{s}},b} \right|^2
     \equiv g^2 | I_{e, {\boldsymbol{s}} } |^2
    \label{eq:excitation-probability-expression}
    \; ,
\end{equation}
$g I_{e, {\boldsymbol{s}} }$ is the corresponding probability amplitude (which, more precisely, is the integral above multiplied by $-i$).
 In a similar way, the absorption probability is given by
\begin{equation}
    P_{a, {\boldsymbol{s}} }
     = \left|\int d\tau \;\bra{0,a}V_I(\tau)\ket{1_{\boldsymbol{s}},b}\right|^2 \equiv g^2 | I_{a, {\boldsymbol{s}} } |^2 
     \label{eq:P_ab_expression}
     \; ,
\end{equation}
 where $g I_{a, {\boldsymbol{s}} }$ is the absorption probability amplitude.
\indent
Equations~(\ref{eq:excitation-probability-expression}) and (\ref{eq:P_ab_expression})
can be evaluated 
with the interaction potential of Eq.~(\ref{eq:interaction-potential}) and using the proper-time parametrized
atom's geodesic (free-fall spacetime trajectory) with given initial conditions. This problem does not have a closed analytical form, but the near-horizon approximation gives expressions for the geodesics and the excitation probability. We will show that, the final result, asymptotically exact in the near-horizon expansion, resembles a Planck distribution.

\section{Near-horizon spacetime trajectory of the atom in free fall}
\label{sec:nh-spacetime-trajectory}
\noindent
The free-fall spacetime trajectory of the atom can be found most efficiently by reducing the geodesic equations to their first-order form, using 
 the four constants of the motion:
the energy $E$ and the axial component of angular momentum $L_z$, given by
\begin{equation}
E= -p_{t} = -   \boldsymbol\xi_{(t)} \cdot {\bf p}
\; , \; \; \; 
L_z= p_{\phi}
  = \boldsymbol\xi_{(\phi)} \cdot {\bf p}
  \; 
  \end{equation}
 (as follows from the 4-momentum $ {\bf p}$ and the 
Killing vectors), in addition to
the invariant mass $\mu$ and the Carter constant $\mathcal{Q}$~\cite{MTW}. 
In this work,
as we will explicitly use the proper time $\tau$ along the geodesic.
(An alternative, convenient choice, e.g., in Ref.~\cite{MTW}, is to use a rescaled affine parameter $\lambda = \tau/\mu $.) 
Moreover, we will rewrite these equations with the
 specific conserved quantities (normalized by mass)
\begin{equation}
e=\frac{E}{\mu}
\; , \; \; \;
 \ell =\frac{ L_z}{\mu}
  \; , \; \;
 \end{equation}
 and
 \begin{equation}
\mathfrak{q}
 = \frac{\mathcal{Q}}{\mu^2}
 =\left(\frac{p_{\theta }}{\mu}\right)^{2}
 +\cos ^{2}\theta \left[a^{2}\left( 1 - e^2 \right)+\left( \frac{\ell}{\sin \theta } \right)^{2}\right] 
\, .
\end{equation}
 The standard procedure yields~\cite{MTW}
\begin{align}
\displaybreak[0]
\rho^2 \frac{dr}{d\tau} &= - \sqrt{\mathcal{R}(r)} \label{eq:geodesic1}\\
\displaybreak[0]
\rho^2 \frac{d\T}{d\tau} &= \pm\sqrt{\varTheta(\T)}\label{eq:geodesic2}\\
\displaybreak[0]
\rho^2 {\frac {d\phi }{d\tau }}
&=-\left(ae-{\frac {\ell}{\sin ^{2}\theta }}\right)+{\frac {a}{\Delta }}\mathcal{P}(r)\label{eq:geodesic3}\\ 
\displaybreak[0]
\rho^2 \frac{dt}{d\tau} &= -a(ae \s^2\T - \ell) + \frac{r^2+a^2}{\D} \mathcal{P}(r) \label{eq:geodesic4}
\; .
\end{align}
In Eqs.~(\ref{eq:geodesic1})--(\ref{eq:geodesic4}), the following auxiliary quantities have been defined
\begin{align}
\displaybreak[0]
\mathcal{P}(r) &= e(r^2+a^2)-a\ell \label{eq:P} \\
\displaybreak[0]
\mathcal{R}(r) &= \left[ \mathcal{P}(r) \right]^2 - \D \left[r^2+(\ell-ae)^2+\mathfrak{q} \right] \label{eq:R} \\
\displaybreak[0]
\varTheta(\T) &= \mathfrak{q}-\C^2\T \left[a^2(1-e^2) + \frac{\ell^2}{\s^2\T}\right]    \label{eq:Theta}  \; .
\end{align}
 The set of equations~(\ref{eq:geodesic1})--(\ref{eq:geodesic4}) gives $dx^{\mu}(\tau)/d \tau$, 
 where $x^{\mu} (\tau) \equiv \bigl(t,  r, \theta, \phi \bigr)$,
 in terms of functions of $(r,\theta)$, independently of $\phi$ and $t$.
This coordinate independence 
 is due to the axisymmetric and stationary nature of the metric. But the combined dependence with respect to 
 $(r,\theta)$ still makes Eqs.~(\ref{eq:geodesic1})--(\ref{eq:geodesic4}) be apparently coupled
 in the form 
$
dx^{\mu} (\tau)/d \tau =
 U^{\mu} (r, \cos \theta)
$.
However, decoupling of this system is possible by combining the corresponding geodesic equations~(\ref{eq:geodesic1}) and (\ref{eq:geodesic2}) into the single separable equation
$
dr/{\sqrt{\mathcal{R}(r)}} = \mp  d \T /{\sqrt{\mathcal{\varTheta (\T)}}}
$,
 where $R(r)$ and $\Theta(\theta)$ are given in Eqs.~(\ref{eq:P})--(\ref{eq:Theta}). 
  With the substitution $y \equiv \C \T $ and writing
$F(y)= 1/\sqrt{\varTheta (y) \, (1-y^2)} $, this integrated equation becomes
\begin{equation}
\int \frac{dr}{\sqrt{\mathcal{R}(r)}} =
 \mp  \int F(y)  {d y} 
\, .
\label{eq:R-Theta_diff-int-elliptic}
\end{equation}
In principle, Eq.~(\ref{eq:R-Theta_diff-int-elliptic}) gives a formal solution $y( r)$ 
in terms of elliptic integrals---the general solutions for all cases are discussed in Ref.~\cite{Chandrasekhar}. Therefore, this separation procedure shows that all the geodesic equations~(\ref{eq:geodesic1})--(\ref{eq:geodesic4}) can be reparametrized as functions of $r$; explicitly,
\begin{equation}
\frac{dx^{\mu} (\tau) }{ d \tau }
= U^{\mu} (r, y (r))
\; .
\label{eq:geodesic-formal}
\end{equation}
The obvious strategy implied by Eq.~(\ref{eq:geodesic-formal}), motivated by the physics (e.g., here for a freely falling system), is to follow the motion with respect to the radial variable $r$ (e.g., as the event horizon is approached).
 Of course, the geodesics are 
  parametrized in terms of the proper time $\tau$, but the radial geodesic~(\ref{eq:geodesic1}) provides a relationship between $\tau$ and $r$ (including the decoupling procedure). Once the inverse relation $\tau (r)$ is formally established, the other geodesic equations~(\ref{eq:geodesic2})--(\ref{eq:geodesic4}) give the complete set of formal 
  solutions $x^{\mu} (\tau) \equiv \bigl( t(r), r, \theta(r), \phi(r) \bigr)$ as functions of $r$.
  In general, this is a difficult problem; however, for our purposes, we will find explicit and remarkably simple near-horizon equations, as will be shown next. 
\\ \indent
 The near-horizon limit of the geodesic equations can be enforced by expanding around  $r_+$, in terms of the variable $x=r-r_+$, using the strategy outlined in the previous paragraph. The near-horizon expansion of the radial geodesic~(\ref{eq:geodesic1}) becomes
 \begin{equation}
 \rho_+^2\frac{dx}{d\tau} 
 \stackrel{(\mathcal H)}{\sim} -\sqrt{c_0^2-c_1x + \mo( x^2) }
\label{eq:geodesic1-nh} 
\, ,
\end{equation}
where $\rho_+^2 \equiv \rho_{+}^2 (\theta) 
= r_+^2 + a^2 \C^2\T$. The constants 
\begin{align}
c_0 &=  
 \mathcal{P}(r_{+}) 
= \left( r_{+}^2 + a^2 \right) \left( e - \Omega_{H} \ell \right) 
 \label{eq:c_0} \\
c_1 &= -4e r_{+} c_{0} 
+ \D'_{+}  \left[ r_+^2+(\ell - a e )^2 +\mathfrak{q} \right]
\label{eq:c_1} 
\, 
\end{align}
are dependent on the conserved quantities of the motion; in particular,
$c_0 $ is proportional to the energy measured in the 
frame dragged with the angular velocity $\Omega_{H}$, i.e.,
\begin{equation}
\tilde{e} = -   \boldsymbol\xi_{(\tilde{t})} \cdot {\bf p}
= e - \Omega_{H} \ell > 0
\; .
\end{equation}
At first sight, the strategy of 
describing the motion in terms of $x$ (i.e., the task of finding the functional relationship $\tau = \tau (x)$)
is complicated by the $\theta$ dependence in the radial geodesic equation~(\ref{eq:geodesic1}) via the quantity $\rho_{+}^2 (\theta)$, a problem that persists in the near-horizon limit~(\ref{eq:geodesic1-nh}). 
But this is the problem whose formal solution we outlined 
with Eqs.~(\ref{eq:R-Theta_diff-int-elliptic}) and (\ref{eq:geodesic-formal}). 
In particular, the near-horizon form of Eq.~(\ref{eq:R-Theta_diff-int-elliptic}) reduces to
\begin{equation}
\int \frac{dx}{ \sqrt{c_0^2-c_1x} }
 \stackrel{(\mathcal H)}{\sim}  \mp  \int F(y)  {d y} 
\, ,
\label{eq:R-Theta_int-nh}
\end{equation}
where the constants are given in Eqs.~(\ref{eq:c_0}) and (\ref{eq:c_1}); again, this is 
expressible in terms of elliptic integrals of the first kind~\cite{Chandrasekhar}. 
Therefore, this procedure shows that all the geodesic equations~(\ref{eq:geodesic1})--(\ref{eq:geodesic4}) can be reparametrized with respect to $x$, as 
implied by Eqs.~(\ref{eq:geodesic-formal}) and (\ref{eq:R-Theta_int-nh}). 
  An exact solution for $y( x) $ can be circumvented in the near-horizon limit as the particle will reach the horizon at a given value $\theta_{+}$ of the polar coordinate. The parameter  $\theta_{+}$ has a simple interpretation: it is an effective ``initial condition'' for the particle to cross the horizon with $\theta$-dependent coordinate values (due to the axisymmetric geometry). This approach involves the leading near-horizon order, so that  $G(\C \theta) \stackrel{(\mathcal H)}{\sim}  G( \C \theta_{+}) $ for any function with nonvanishing zeroth order. Specifically, this leads to the replacement $\rho^2  \stackrel{(\mathcal H)}{\sim} \hat{\rho}_{+}^2 \equiv r_{+}^2 + a^2 \cos^{2}\theta_{+}$.
 In other words, for the near-horizon radiation problem, the relevant range of $\theta$  can be restricted. 
Then, the proper-time relation $\tau = \tau (x)$ follows by integrating Eq.~(\ref{eq:geodesic1-nh}), 
\begin{equation}
\tau = -k  x+ \mo(x^2) + \text{const} \, ,
\label{eq:tau-vs-x}
\end{equation}
where $k=\hat{\rho}_+^2/c_0$. Next, to obtain the functional relationships $t = t(x)$ and $\phi= \phi(x)$ to leading order,
we can divide Eqs.~(\ref{eq:geodesic3}) and (\ref{eq:geodesic4}) respectively by Eq.~(\ref{eq:geodesic1})---
this directly removes the $\theta$-dependent $\rho^2$ factors; 
thus, by straightforward integration, the near-horizon expansions are
\begin{align}
    t  &\stackrel{(\mathcal H)}{\sim} - \frac{1}{2 \kappa} \ln x - C x +\mo(x^2) \; , \label{eq:t-vs-x} \\
    \tilde{\phi} &\stackrel{(\mathcal H)}{\sim} \alpha x +\mo(x^2)\; .  \label{eq:phi-vs-x}
\end{align}
In Eq.~(\ref{eq:phi-vs-x}),
we have evaluated $\tilde{\phi} = \phi-\Omega_H t$ (by combining solutions for $\phi$ and $t$) because the CQM modes in Eq.~(\ref{eq:CQM_modes}) explicitly depend on this corotating azimuthal variable. Most importantly, even though both $\phi$ and $t$ have logarithmic terms proportional to $\ln x$, these cancel out when combined into the locally well-defined coordinate $\tilde{\phi}$.
Finally, for the sake of completeness, the constants $C$ and $\alpha$ can be computed by collecting all the $\mathcal{O}(x)$ terms arising from the functions on the right-hand side of Eqs.~(\ref{eq:geodesic1})--(\ref{eq:geodesic4}); a straightforward  calculation gives
\begin{equation}
C = \frac{1}{2 \kappa} 
\left[ \frac{1}{2} \frac{c_{1}}{c_{0}^{2}} + 
\frac{2 r_{+}}{R_{+}^2} \frac{\left( \tilde{e} + \Omega_H \ell \right) }{\tilde{e}}
- \frac{1}{2 \kappa R_{+}^{2}}
- \frac{\Omega_{H}}{\tilde{e}} \left( a e s_{+}^2 - \ell \right)
\right]
\label{eq:constant-C_Kerr}
\end{equation}
and
\begin{equation}
\alpha
=
\Omega_{H} \frac{r_{+}}{\kappa R_{+}^2}
- \frac{ \left( a e s_{+}^2 - \ell \right) \left( \Omega_{H} a - 1/ s_{+}^2 \right) }{R_{+}^2 \tilde{e}}
\; ,
\label{eq:constant-alpha}
\end{equation}
where $R_{+}^2 = r_{+}^2 + a^2$ and $s_{+} = \sin \theta_{+}$.
Remarkably, the constants $C$, $\alpha$, and $k$, as shown in Sec.~\ref{sec:nh-excitation-probability}, do not play a direct role in the radiation formula.

\section{Near-horizon excitation probability and radiation properties}
\label{sec:nh-excitation-probability}
\noindent
We are now ready to use the reparametrization of the geodesics in terms of the near-horizon coordinate $x$ to compute the emission probability for the atom in free fall.
Replacing Eqs.~(\ref{eq:CQM_modes}), (\ref{eq:tau-vs-x}), (\ref{eq:t-vs-x}), and $\tilde{\phi} (x)$ 
in Eq.~(\ref{eq:excitation-probability-expression}) gives
\begin{equation}
P_{e, {\boldsymbol{s}} }  
    = g^2 k^2 \left|\int_0^{x_f} dx \;x^{-i \sigma} e^{-i q x} \right|^2\;,
    \label{eq:near-horizon-pex}
\end{equation}
where $x_f$ is an upper bound of the near-horizon approximation, $k=\hat{\rho}_+^2/c_0$, and the parameters in the integral are
$\sigma = \tilde{\omega}/\kappa$ (for the purely outgoing radiation modes $\Phi^{ + {\rm \scriptscriptstyle (CQM)} }_{\boldsymbol{s}} $)
and 
$q=C\tilde{\omega}+k \nu + \alpha m $.
 The integrand of Eq.~(\ref{eq:near-horizon-pex}) involves two competing oscillatory functions 
 $f_1(x)\equiv x^{-i\tilde{\omega}/\kappa} = e^{-i (\tilde{\omega}/\kappa)\ln x} 
$ and $f_2(x)\equiv e^{-iqx}$ that select the near-horizon region~\cite{camblong2020}, with
$\nu \gg \tilde{\omega} \, , \alpha m $, and the outcome is independent of $C$ and $\alpha$. Such behavior displays the dominant physics that is invariant under arbitrary magnifications. This form of conformal dominance is due to the CQM modes and the associated logarithmic dependence of $t$ on $x$, with the near-horizon wavefronts piling up in a Russian-doll geometric sequence~\cite{camblong2020}. 
 Then, the amplitude integral is given by 
 $ \int_0^{x_f} dx\; x^{-i\sigma} e^{ -i q x  } 
  \approx  {q}^{-1}
  \sqrt{  2 \pi \sigma /  (e^{2 \pi \sigma} -1) }
    \; e^{i \delta} $, where $\delta$ is a real phase. 
 Thus, the resulting 
  radiation spectrum is governed by the emission probability
\begin{equation}
    P_{e, {\boldsymbol{s}} }   = \frac{2\pi g^2 \tilde{\omega}}{\kappa \nu^2} \; 
    \left( e^{2\pi\tilde{\omega}/\kappa}-1 \right)^{-1}
    \, ,
    \label{eq:p-ex-planck}
\end{equation}
which is the central result of this paper.
\\ \indent
Some important remarks are in order.
First, the only nonzero contributions in Eq.~(\ref{eq:near-horizon-pex}), leading to Eq.~(\ref{eq:p-ex-planck}),
 arise from the purely outgoing CQM component
$\Phi^{ + {\rm \scriptscriptstyle (CQM)} }_{\boldsymbol{s}} $, for which
$\sigma = \tilde{\omega}/\kappa$; 
if, instead, the ingoing CQM component were used, the logarithmic terms would cancel, yielding $\sigma= 0$ and a vanishing outcome.
 For that reason, {\em any generic Boulware-like state $\left| B \right\rangle$ will give a Planck distribution~(\ref{eq:p-ex-planck})\/}. (See Appendix~\ref{sec:field-modes-vacuum}.)
\\ \indent
Second,
it is noteworthy that the probability amplitude of Eq.~(\ref{eq:p-ex-planck}) corresponds to a Planck statistical distribution that is a function of the variable $\tilde{\omega}=\omega-m\Omega_H$, with $m\Omega_H$ as a generalized chemical potential that favors the black hole's tendency to remove its conserved quantum numbers. Incidentally, this functional combination corresponds to the thermodynamic change $\delta M - \Omega_{H} \delta J$, which also relates to the coupling with fields and particles (orbital parameters, gyroscope precession, and Sagnac effect~\cite{Scully-metric-test}).
\\ \indent
Finally, the appearance of the Planck function in Eq.~(\ref{eq:p-ex-planck}) shows its apparent equivalence to a thermal distribution with the Hawking temperature $T_{H}= \beta^{-1}_{H}=\kappa/2\pi$ (proportional to the surface gravity $\kappa$). 
Interestingly, this arises from the emission of a pure state by a single atom, with definite correlations between the modes encoded in the phase of the integral in 
Eqs.~(\ref{eq:excitation-probability-expression}) and (\ref{eq:near-horizon-pex}).
However, the setup can be extended to a model consisting of an ensemble of freely-falling atoms forming a cloud, as in Refs.~\cite{scully2018,scully2019}; if the initial conditions for their spacetime trajectories are random, then, the outgoing radiation field would be {\em effectively thermal\/}. 
Mathematically,  
Eqs.~(\ref{eq:excitation-probability-expression}) and (\ref{eq:p-ex-planck}) give the probability of emission of a field quantum; 
 similarly, Eq.~(\ref{eq:P_ab_expression}) gives
   the absorption probability  $P_{a, {\boldsymbol{s}} } $ 
 (for  the transition from field state
${1_{\boldsymbol{s}}}$ to
${0_{\boldsymbol{s}}}$)~\cite{QED-accel-Scully-etal_1,QED-accel-Scully-etal_2},
 which formally reduces to $\tilde{\omega} \rightarrow - \tilde{\omega}$;
this yields 
\begin{equation}
\frac{P_{e, {\boldsymbol{s}} }   }{ P_{a, {\boldsymbol{s}} }  }
 = e^{-\beta_{H} \tilde{\omega}}
 \; .
 \label{eq:ratio_em_ab_Boltzmann}
 \end{equation}
The interpretation of the ratio~(\ref{eq:ratio_em_ab_Boltzmann}) as modeling a thermal distribution 
with a Boltzmann factor~\cite{QED-accel-Scully-etal_1,QED-accel-Scully-etal_2}
has been used for black hole thermodynamics~\cite{Boltzmann-BH_1, Boltzmann-BH_2}. Furthermore, 
the physical origin of this factor can be traced to the CQM waves, as in Eq.~(\ref{eq:near-horizon-pex}).
The Boltzmann-factor analysis of the thermal nature of the radiation
can be expanded by considering the reduced density matrix 
of the field, as in Ref.~\cite{scully2018},
where the radiation field is called horizon brightened acceleration radiation (HBAR).
The corresponding master equation for the diagonal elements ${\rho}_{n,n} $ of a given single mode,
\begin{equation}
    \dot{\rho}_{n,n} =
    - R_{{\mathrm e}, {\boldsymbol{s}} } \big[(n+1) \, {\rho}_{n,n} - n \,  {\rho}_{n-1,n-1}\big] -  
    R_{{\mathrm a}, {\boldsymbol{s}} }\big[n \, {\rho}_{n,n}-(n+1) \, {\rho}_{n+1,n+1} \big] 
    \; ,
  \label{eq:master_equation_final_single-mode}
  \end{equation}
     admits a steady-state distribution that is indeed thermal for random initial injection times of the atomic cloud. 
     Here, the rate coefficients $R_{{\mathrm e}, {\boldsymbol{s}} }$ and $R_{{\mathrm a}, {\boldsymbol{s}} }$
   are proportional  to $P_{{\mathrm e}, {\boldsymbol{s}} }$ and $P_{{\mathrm a}, {\boldsymbol{s}} }$.
     The implementation of the thermal condition of the radiation field requires
     that the Boltzmann factor~(\ref{eq:ratio_em_ab_Boltzmann}) be satisfied as above for all field modes.    
     A detailed analysis of this HBAR density-matrix approach---including nontrivial generalizations to simultaneous modes, to the Kerr geometry, and with an all-encompassing thermodynamic correspondence---is in progress, and will be reported elsewhere.

\section{Discussion}
\label{sec:discussion}
\noindent
In this paper, we have shown that the acceleration radiation of an atom falling freely into a Kerr black hole in a Boulware-like vacuum is driven by the near-horizon physics. 
Specifically, this Unruh (acceleration) radiation can be traced to the dominance of near-horizon CQM modes in the excitation probability.
This radiation is due to the acceleration of the reference frame in which the vacuum field modes are defined with respect to 
the freely falling atom (locally inertial frame)~\cite{UA-mirror-Svidzinsky}; thus,
its existence agrees with the qualitative equivalence principle~\cite{fulling2018},
and generates a spectrum with the Hawking-Unruh temperature $T=\kappa/2\pi$.
Furthermore, the HBAR radiation field of an atomic cloud with random initial times has a thermal character, as will be further analyzed in a forthcoming article.
\\ \indent
Moreover, our analysis covers all 4D black holes subject to the no-hair theorem~\cite{MTW}:  the Schwarzschild geometry ($a=0$, as in Ref.~\cite{camblong2020}) and Kerr-Newman black holes with electric charge $Q$, obtainable via
$\D \longrightarrow  r^2 -2Mr + a^2 + Q^2$ (in geometrized units, with unit Coulomb constant). This shows the {\it universality\/} of this form of conformal symmetry and the {\it robustness\/} of the ensuing acceleration radiation for all 4D black holes.
\\ \indent
Finally, our work highlights the simplicity of the near-horizon framework as a tool to tackle otherwise intractable problems, and elucidates the connection of the acceleration radiation with (0+1)-dimensional CFT. 

\acknowledgments{}
M.O.S.\ and A.A.\ acknowledge support by the
National Science Foundation (Grant No.\ PHY-2013771), the Air Force Office of Scientific Research (Award No.\ FA9550-20-1-0366 DEF), the Office of Naval Research (Award No.\ N00014-20-1-2184), the Robert A. Welch Foundation (Grant No.\ A-1261) and the King Abdulaziz City for Science and Technology (KACST).
H.E.C. acknowledges support
by the University of San Francisco Faculty Development Fund. 
This material is based upon work supported by the Air Force Office of Scientific Research under award FA9550-21-1-0017 (C.R.O. and A.C.).

\begin{appendix}

\section{Separation of variables and derivation of the near-horizon CQM equation} 
\label{app:CQM_derivation}
\noindent
In this appendix, we summarize two important topics: (i) the basics of separation of variables in Boyer-Lindquist coordinates; and (ii) three derivations of the near-horizon CQM equation.

\subsection{Separation of variables in Boyer-Lindquist coordinates}
When the separation of variables of Eq.~(\ref{eq:KG_separation}) is enforced
in Eq.~(\ref{eq:Kerr_Klein_Gordon_spatial}),
the radial function $R(r)$ satisfies the  equation
\begin{equation}
 \frac{d}{d r} \left( \D  \frac{d R}{d r} \right) +
\left[\frac{(r^2+a^2)^2\omega^2 - 4 Mra m \omega + a^2 m^2}{\D} - \Lambda_{{\boldsymbol{s}}} - a^2 \omega^2 - \mu_\Phi^2 r^2
\right] R = 0
\label{eq:Kerr_Klein_Gordon_radial}
\, ,
\end{equation}
where $ \Lambda_{{\boldsymbol{s}}} $ is the separation constant extracted from the angular equation for spheroidal wave functions~\cite{spheroidal,frolov}
\begin{equation}
\frac{1}{\s \T}  \frac{d}{d \T} \left( \s \T  \frac{d S}{d \T} \right) 
+
\left[ a^2 \omega^2 \cos^2 \theta - \frac{m^2}{\sin^2 \theta} 
+  \Lambda_{{\boldsymbol{s}}}
-
a^2 \mu_\Phi^2  \cos^2 \theta \right]
 S = 0
 \label{eq:Kerr_Klein_Gordon_polar}
\, 
\end{equation}
(most commonly studied for $ \mu_\Phi=0$).
As Eq.~(\ref{eq:Kerr_Klein_Gordon_polar}) defines a Sturm-Liouville problem, the regular solutions 
form a complete orthogonal set labeled by a discrete ``spheroidal'' number $l$.
Additionally, the
 separation constant $\Lambda_{{\boldsymbol{s}}}$ depends on the other quantization parameters: $m$
 and $\omega$, as well as the Kerr parameter $a$ and the field mass $\mu_\Phi$
 (both $\omega$ and $\mu_\Phi$
appear in the dimensionless combinations $a \omega $ and  $ a \mu_\Phi $).

\subsection{Derivation of the near-horizon CQM equation}
A first derivation starts from the full-fledged radial equation~(\ref{eq:Kerr_Klein_Gordon_radial}). Implementing the near-horizon expansion with Eq.~(\ref{eq:n-h_expansion})
selects the dominant terms, 
which are the first one (with radial derivatives) and the ratio in the square brackets.
 In the latter, it is straightforward to complete squares  using the definition of the angular velocity $\Omega_H$ of the black hole. This converts the radial equation (\ref{eq:Kerr_Klein_Gordon_radial}) into the leading radial near-horizon equation
\begin{equation}
\left[\frac{1}{x}\frac{d}{d  x} \left( x  \frac{d}{d x} \right)+\left( \tilde{\omega} \frac{(r^2+a^2)}{\D_{+}'} \right)^2
\frac{1}{x^2} \right] R(x) \stackrel{(\mathcal H)}{\sim} 0 \; ,
\label{eq:Kerr_Klein_Gordon_conformal-R_app}
\end{equation}
which is equivalent to Eq.~(\ref{eq:Kerr_Klein_Gordon_conformal-R}), when
Eq.~(\ref{eq:surface-gravity_Kerr}) for the surface gravity is used.
As shown in the main text, Eq.~(\ref{eq:Kerr_Klein_Gordon_conformal-R_app}) can be reduced to its normal form with the Liouville transformation that yields the standard CQM Hamiltonian of Eq.~(\ref{eq:Kerr_Klein_Gordon_conformal}).

A second derivation involves the alternative, equivalent expression for the Kerr metric given by Eq.~(\ref{eq:Kerr3}).
While this equation gives the covariant metric, it can easily be inverted to get the contravariant components needed for the Klein-Gordon equation~(\ref{eq:Klein_Gordon_basic}). 
Then, in the near-horizon region,
instead of Eq.~(\ref{eq:Kerr_Klein_Gordon_spatial}) or
Eq.~(\ref{eq:Kerr_Klein_Gordon_radial}), one can directly write
\begin{align}
&
\left[
- \frac{\Sigma^2}{\rho^2 \Delta} \frac{\partial}{\partial  \tilde{t}^2}
+
\frac{\rho^2 }{\Sigma^2 \s^2 \T } \frac{\partial}{\partial  \tilde{\phi}^2}
+
\frac{1}{\rho^2} \frac{\partial}{\partial  r} \left( \D  \frac{\partial}{\partial r} \right)
+
\frac{1}{\rho^2} \frac{\partial}{\partial  \T^2}
\right]
\Phi
\\
&
\stackrel{(\mathcal H)}{\sim}
\left[
- \frac{(r^2+a^2)^2}{\rho^2 \Delta} \frac{\partial}{\partial  \tilde{t}^2}
+
\frac{1}{\rho^2} \frac{\partial}{\partial  r} \left( \D  \frac{\partial}{\partial r} \right)
\right] \Phi
\stackrel{(\mathcal H)}{\sim}
0
\; ,
\label{eq:Kerr_Klein_Gordon_conformal-with-time}
\end{align}
due to the leading behavior 
$\D(r)  \stackrel{(\mathcal H)}{\sim}  \D'_{+}  \, x$, which  selects the radial-time sector of the metric.
Equation~(\ref{eq:Kerr_Klein_Gordon_conformal-with-time})
reproduces again the asymptotically exact equation~(\ref{eq:Kerr_Klein_Gordon_conformal-R}).

A third derivation can be completed by using the equivalent tortoise coordinate $r_*$ for the Kerr metric, which is defined through the equation 
 \begin{equation}
\frac{dr_*}{dr} = 
\frac{1}{f(r)}
\label{eq:tortoisedef}
\; ,
\end{equation}
 where $ f\equiv {\D} /({r^2+a^2})$, so that
\begin{equation}
    r_* = \int \frac{r^2+a^2}{\D} dr 
    \; .
    \label{eq:tortoise_r}
\end{equation}
This coordinate choice is made so that the radial-time sector of the metric appears as 
near-horizon conformally flat and pushes the horizon radially to minus infinity. Notice that the scale factor $f(r)$ plays the same role as the homologous factor in generalized Schwarzschild coordinates.
In the corotating coordinates~(\ref{eq:rotating-coords_separation_1}), the radial function $R(r)$ satisfies the wave equation
\begin{equation}
\left[\frac{d^2}{dr_*^2}+\tilde{\omega}^2\right] R(r)=0
\label{eq:Kerr_conformal-tortoise}
\, .
\end{equation}
Most importantly,
Eq.~(\ref{eq:Kerr_conformal-tortoise}) with the tortoise coordinate is equivalent to its counterpart
with the regular Boyer-Lindquist radial variable, Eq.~(\ref{eq:Kerr_Klein_Gordon_conformal-R_app})
or Eq.~(\ref{eq:Kerr_Klein_Gordon_conformal-R}).
The ingoing and outgoing waves 
$\left\{ e^{-i\tilde{\omega}(\tilde{t}+r_*)},e^{-i\tilde{\omega}(\tilde{t}-r_*)}\right\}$ in terms of $r_*$ 
correspond to the conformal ingoing/outgoing modes $x^{\mp i \Theta}$ of CQM.
This is expected from the fact that we have just mapped the near-horizon physics from one coordinate frame to another.
A simple proof of this equivalence, at the level of the differential equations,
follows from the definition~(\ref{eq:tortoisedef}) of this coordinate transformation, 
whose near-horizon leading form $dr_*/dr \stackrel{(\mathcal H)}{\sim}dr_*/dx 
\stackrel{(\mathcal H)}{\sim} 1/(f'_{+} x )$ implies that
\begin{equation}
 \frac{d^2}{dr_*^2} \stackrel{(\mathcal H)}{\sim} 
(f_+^{\prime})^2 \, x \frac{d}{d  x} \left( x  \frac{d}{d x} \right)
 \Longrightarrow 
 \left[ \frac{1}{x} \frac{d}{d  x} \left( x  \frac{d}{d x} \right)
+\frac{\tilde{\omega}^2}{(f_+^{\prime})^2}\right] R (x) \stackrel{(\mathcal H)}{\sim} 0
\, ,
\end{equation}
where $f'_{+} = \Delta'_{+} /(r_{+}^2+a^2)$,
which is identical to Eq.~(\ref{eq:Kerr_Klein_Gordon_conformal-R}).

\section{Field modes and vacuum states}
\label{sec:field-modes-vacuum}
\noindent
Kerr field modes and vacuum states have been analyzed in Refs.~\cite{Kerr-vacuum-states_1,Kerr-vacuum-states_2}.
A vacuum state $\left| 0 \right\rangle$
is defined via 
 a complete set of positive-frequency modes
$\Phi_{\boldsymbol{s}}$ with annihilation operators $ \hat{a}_{\boldsymbol{s}}$
such that  $ \hat{a}_{\boldsymbol{s}} \left| 0 \right\rangle = 0$. The Kerr-geometry modes below are denoted 
$\varphi^{\Lambda}_{\boldsymbol{s}}$, where $\boldsymbol{s}  = \left\{ \omega, l, m\right\}$ and
  $\Lambda$  labels  {\em in, up, out,} and {\em down\/}~\cite{Chrzanowski}.
With the notation $\mathcal{H}^{\mp}$ for the past/future horizons  
and $\mathscr{I}^{\mp}$  for past/future null infinity,
the up mode $\varphi^{\mathrm{up}}_{\boldsymbol{s}} $ (with radial function $R^{-}_{\boldsymbol{s}}$)
initially emerges from $\mathcal{H}^{-}$ (reflecting back to $\mathcal{H}^{+}$ and transmitting to $\mathscr{I}^{+}$, with coefficients $A^{-}_{\boldsymbol{s}}$ and $B^{-}_{\boldsymbol{s}}$); 
and the in mode $\varphi^{\mathrm{in}}_{\boldsymbol{s}} $ (with radial function $R^{+}_{\boldsymbol{s}}$)
is initially ingoing from $\mathscr{I}^{-}$ (reflecting and transmitting to $\mathscr{I}^{+}$ and
$\mathcal{H}^{+ }$ respectively, with coefficients $A^{+}_{\boldsymbol{s}}$ and $B^{+}_{\boldsymbol{s}}$). The out and down modes are the corresponding time-reversed solutions 
(i.e., $R^{\mathrm{out}}_{\boldsymbol{s}}= R^{+ *}_{\boldsymbol{s}}$
and  $R^{\mathrm{down}}_{\boldsymbol{s}}= R^{- *}_{\boldsymbol{s}}$).
Moreover, these modes map asymptotically to the CQM modes~(\ref{eq:CQM_modes})
 in the form
\begin{equation}
\varphi^{\mathrm{out}}_{\boldsymbol{s}} 
\stackrel{(\mathcal H)}{\propto} 
\Phi^{ + {\rm \scriptscriptstyle (CQM)} }_{\boldsymbol{s}} 
\;  , \; \; \;  \; \; \; 
\varphi^{\mathrm{in}}_{\boldsymbol{s}} 
\stackrel{(\mathcal H)}{\propto} 
\Phi^{ - {\rm \scriptscriptstyle (CQM)} }_{\boldsymbol{s}} 
\label{eq:CQM-vs-global-modes}
\end{equation}
(with the proportionality coefficients defined above; also
$
R^{\mp}_{\boldsymbol{s}} \stackrel{(\mathcal H)}{\propto}  R^{ \pm {\rm \scriptscriptstyle (CQM)}}
$).
By contrast, the up and down modes $\varphi^{\mathrm{up, down}}_{\boldsymbol{s}} $ include both near-horizon components 
$\Phi^{ \pm {\rm \scriptscriptstyle (CQM)} }_{\boldsymbol{s}} $.

Now, generic Boulware-like vacua are defined via the positive-frequency modes above with respect to the Killing vector $\boldsymbol\xi_{({t})} = \partial/\partial t$, i.e., in Boyer-Lindquist coordinates.
The past Boulware vacuum $\left| B^{-} \right\rangle$ is defined in terms of the basis
$\left\{ \varphi^{\mathrm{up}}_{\boldsymbol{s}} , \varphi^{\mathrm{in}}_{\boldsymbol{s}} \right\}$ (with values on the past Cauchy surface $\mathcal{H}^{-} \cup \mathscr{I}^{-}$); and similarly for the future Boulware vacuum 
$\left| B^{+} \right\rangle$ in terms of 
$\left\{ \varphi^{\mathrm{out}}_{\boldsymbol{s}}  ,\varphi^{\mathrm{down}}_{\boldsymbol{s}} \right\}$
(with Cauchy values on  $\mathcal{H}^{+} \cup \mathscr{I}^{+}$).
In Ref.~\cite{Kerr-vacuum-states_1}, $\left| B^{-} \right\rangle$ and $\left| B^{+} \right\rangle$
were shown to be inequivalent due to the superradiant modes needed for a complete basis---these are the up/down modes (in each basis) with the frequency range 
$ 0< \omega <  m \Omega_{H} $ 
(for co-rotating waves, $m>0$).
This problem originates from the mismatch of frequencies $\omega$ and $\tilde{\omega}$ associated 
with the Killing vectors $\boldsymbol\xi_{({t})} $ and $\boldsymbol\xi_{(\tilde{t})}$
(naturally adapted to $ \mathscr{I}^{\pm}$  and $\mathcal{H}^{\pm} $ respectively).
We will now use these Boulware states and Eq.~(\ref{eq:CQM-vs-global-modes}) to find the excitation probability of a freely falling atom.  

As discussed in Sec.~\ref{sec:nh-excitation-probability},
 {\em any generic Boulware-like state $\left| B \right\rangle$ will give a Planck distribution~(\ref{eq:p-ex-planck})\/}.
In particular, for the basis sets associated with $\left| B^{\pm} \right\rangle$, the purely outgoing components that give Eq.~(\ref{eq:p-ex-planck}) are extracted accordingly ($\varphi^{\mathrm{out}}_{\boldsymbol{s}} $ for $\left| B^{+} \right\rangle$ and
$ \varphi^{\mathrm{up}}_{\boldsymbol{s}} $ for $\left| B^{-} \right\rangle$); this
leaves the interpretation of the superradiant modes (for $- m \Omega_{H} < \tilde{\omega} < 0$),
 as an additional technicality. 
Moreover, this calculation shows that, for the future Boulware vacuum $\left| B^{+} \right\rangle$, the superradiant modes are subsumed in the Planck distribution~\cite{frolov};
 in addition, from
$ 
\varphi^{\mathrm{out}}_{\boldsymbol{s}} 
\stackrel{(\mathcal H)}{\sim} 
B_{{\boldsymbol{s}}}^{+*}
\Phi^{ + {\rm \scriptscriptstyle (CQM)} }_{\boldsymbol{s}} 
$
[including the proportionality coefficients in
Eq.~(\ref{eq:CQM-vs-global-modes})],
and as  $\mathcal{T}_{{\boldsymbol{s}}} = B_{{\boldsymbol{s}}}^{+*}$
is the amplitude transmission coefficient, the modified
Eq.~(\ref{eq:p-ex-planck}) accounts for the greybody factors
 $\Gamma_{{\boldsymbol{s}}} = |\mathcal{T}_{{\boldsymbol{s}}}|^2$.
If the past Boulware vacuum $\left| B^{-} \right\rangle$ is chosen, the classical superradiant modes also give rise to  the Unruh-Starobinsky radiation (quantum superradiance)~\cite{Unruh-Starobinsky-rad} of the vacuum at  $\mathscr{I}^{+}$.

 Additionally, for fermion fields, the definition of quantum states is less constrained, and there are other candidate Boulware states~\cite{Kerr-fermion-Q}.

\end{appendix}


\begin{thebibliography}{}
\bibitem{hawking1974}
S. W. Hawking, Black hole explosions?, Nature (London) {\bf 248}, 30 (1974).

\bibitem{hawking1975}
S. W. Hawking, Particle creation by black holes, Commun. Math. Phys. {\bf 43}, 199 (1975).

\bibitem{Bekenstein_entropy}
J. D. Bekenstein, Black holes and the second law, Lett. Nuovo Cimento {\bf 4}, 737 (1972);
Black holes and entropy, Phys. Rev. D {\bf 7}, 2333 (1973);
Generalized second law of thermodynamics in black-hole physics, Phys. Rev. D {\bf 9}, 3292 (1974).

\bibitem{hawking_BHmech}
J. M. Bardeen, B. Carter, and S. W. Hawking, The four laws of black hole mechanics, Commun. Math. Phys. {\bf 31}, 161 (1973).

\bibitem{unruh76}
W. G. Unruh, Notes on black-hole evaporation, Phys. Rev. D {\bf 14}, 870 (1976).

\bibitem{fulling73}
S. A. Fulling, Nonuniqueness of canonical field quantization in Riemannian space-time, Phys. Rev. D {\bf 7}, 2850 (1973).

\bibitem{davies74}
P. C. Davies, Scalar production in Schwarzschild and Rindler metrics, J. Phys. A: Math. Gen. {\bf 8}, 609 (1975).

\bibitem{scully2018}
M. O. Scully, S. Fulling, D. M. Lee, D. N. Page, W. P. Schleich, and A. A. Svidzinsky, Quantum optics approach to radiation from atoms falling into a black hole, Proc. Nat. Acad. Sci. USA {\bf 115}, 8131 (2018).

\bibitem{scully2019}
J. S. Ben-Benjamin {\em et al.\/}, Unruh Acceleration Radiation Revisited, Int. J. Mod. Phys. A {\bf 34}, 1941005 (2019).

\bibitem{Chakraborty-Majhi_2019}
K. Chakraborty and B. R. Majhi,
 Detector response along null geodesics in black hole spacetimes and in a Friedmann-Lemaitre-Robertson-Walker universe,
 Phys. Rev. D {\bf 100}, 045004 (2019).
 
\bibitem{Dalui-Majhi-Mishra_2020}
 S. Dalui, B. R. Majhi, and P. Mishra,
Horizon induces instability locally and creates quantum thermality
Phys. Rev. D {\bf 102}, 044006 (2020); 
S. Dalui and B. R. Majhi, 
Near horizon local instability and quantum thermality,
arXiv:2007.14312 [gr-qc].	

\bibitem{camblong2020}
H. E. Camblong, A. Chakraborty, and C. R. Ord\'o\~nez, Near-horizon aspects of acceleration radiation by free fall of an atom into a black hole, 
Phys. Rev. D {\bf 102}, 085010 (2020).

\bibitem{LIGO:2016}
B. P. Abbott {\it et al.\/}, 
 Observation of Gravitational Waves from a Binary Black Hole Merger, Phys. Rev. Lett. {\bf 116}, 061102 (2016).

\bibitem{carlip1}
S. Carlip, Black hole entropy from conformal field theory in any dimension, Phys. Rev. Lett. {\bf 82}, 2828 (1999).

\bibitem{carlip2}
S. Carlip, Symmetries, horizons, and black hole entropy, Gen. Relativ. Gravit. {\bf 39}, 1519 (2007).

\bibitem{strominger}
A. Strominger, Black hole entropy from near-horizon microstates, J. High Energy Phys. 02 (1998) 009.

\bibitem{stromingerkerr}
M. Guica, T. Hartman, W. Song, and A. Strominger, The Kerr/CFT correspondence, Phys. Rev. D {\bf 80}, 124008 (2009).

\bibitem{larsen}
A. Castro and F. Larsen, Near-extremal Kerr entropy from AdS2 quantum gravity, J. High Energy Phys. 12 (2009) 037.

\bibitem{hartman}
T. Hartman, W. Song, and A. Strominger, Holographic derivation of Kerr-Newman scattering amplitudes for general charge and spin, J. High Energy Phys. 03 (2010) 118.

\bibitem{strominger10}
A. Castro, A. Maloney, and A. Strominger, Hidden conformal symmetry of the Kerr black hole, Phys. Rev. D {\bf 82}, 024008 (2010).

\bibitem{parkerprl}
I. Agullo, J. Navarro-Salas, G. J. Olmo, and L. Parker, Hawking radiation by Kerr black holes and conformal symmetry, Phys. Rev. Lett. {\bf 105}, 211305 (2010).

\bibitem{Padmanabhan_BH-ISP}
K. Srinivasan and T. Padmanabhan, Particle Production and Complex Path Analysis, 
Phys. Rev. D {\bf 60}, 24007 (1999);
T. Padmanabhan,
Gravity and the Thermodynamics of Horizons, 
Phys. Rep. {\bf 406}, 49 (2005).

\bibitem{guptasen}
D. Birmingham, K. S. Gupta, and S. Sen, Near-horizon conformal structure of black holes, Physics Letters B, 505(1-4), 191-196 (2001).

\bibitem{nhthermocqm}
H. E. Camblong and C. R. Ord\'o\~nez, Black hole thermodynamics from near-horizon conformal quantum mechanics, Phys. Rev. D {\bf 71}, 104029 (2005).

\bibitem{nhcamblong-sc}
H. E. Camblong and C. R. Ord\'o\~nez, Semiclassical methods in curved spacetime and black hole thermodynamics, Phys. Rev. D {\bf 71}, 124040 (2005).

\bibitem{vaidya}
T. R. Govindarajan, V. Suneeta, and S. Vaidya, Horizon states for AdS black holes, Nucl. Phys. B {\bf 583}, 291 (2000).

\bibitem{moretti}
V. Moretti and N. Pinamonti, Aspects of hidden and manifest SL(2, R) symmetry in 2D near-horizon black-hole backgrounds, Nucl. Phys. B {\bf 647}, 131 (2002).

\bibitem{dff}
V. de Alfaro, S. Fubini, and G. Furlan, Conformal invariance in quantum mechanics, Nuovo Cim. A {\bf 34}, 569 (1976).

\bibitem{HEC-CRO_CQM}
H. E. Camblong and C. R. Ord\'o\~nez, Renormalization in conformal quantum mechanics, Phys. Lett. A {\bf 345}, 22 (2005).

\bibitem{tightnesscamblong}
H. E. Camblong and C. R. Ord\'o\~nez, Conformal tightness of holographic scaling in black hole thermodynamics, Classical and Quantum Gravity {\bf 30}, 175007 (2013).

\bibitem{QED-accel-Scully-etal_1}
M. O. Scully, V. V. Kocharovsky, A. Belyanin, E. Fry, and F. Capasso, Enhancing acceleration radiation from ground-state atoms via cavity quantum electrodynamics, Phy. Rev. Lett. {\bf 91}, 243004 (2003).

\bibitem{QED-accel-Scully-etal_2}
A. Belyanin, V. V. Kocharovsky, F. Capasso, E. Fry, M. S. Zubairy, and M. O. Scully, Quantum electrodynamics of accelerated atoms in free space and in cavities, Phys. Rev. A, {\bf 74}, 023807 (2006).

\bibitem{Landau-Lifshitz_CTFs}
L. Landau and E. M. Lifshitz, The Classical Theory of Fields, 4th ed.\/ (Pergamon Press, Oxford, 1975).

\bibitem{frolov}
V.~P. Frolov and I.~D. Novikov,
{\em Black Hole Physics: Basic Concepts and New Developments\/}
(Springer, 2012).

\bibitem{spheroidal}
C. Flammer, Spheroidal wave functions (Stanford University Press, Stanford, 1957); 
M. Abramowitz and I. A. Stegun, eds., Handbook of Mathematical Functions
(Dover Publications, New York, 1972), Chap.~21.

\bibitem{rotating-Qvacuum}
P. C. W. Davies, T. Dray, and C. A. Manogue,
Detecting the rotating quantum vacuum,
Phys. Rev. D {\bf 53}, 4382 (1996);
N. Nicolaevici,
Null response of uniformly rotating Unruh detectors in bounded regions,
Class. Quantum Grav. {\bf 18}, 5407 (2001).

\bibitem{mirror-no-superradiance}
A. L. Matacz, P. C. W. Davies, and A. C. Ottewill, 
Quantum vacuum instability near rotating stars,
Phys. Rev. D {\bf 47}, 1557 (1993);
G. Duffy and A. C. Ottewill, 
Rotating quantum thermal distribution,
Phys. Rev. D {\bf 67}, 044002 (2003);
G. Duffy and A. C. Ottewill, 
Renormalized stress tensor in Kerr space-time: Numerical results for the
Hartle-Hawking vacuum,
Phys. Rev. D {\bf 77}, 024007 (2008).

\bibitem{MTW}
C. W. Misner, K. S.Thorne, and J. A. Wheeler, Gravitation (W. H. Freeman, San Francisco, 1973).

\bibitem{Chandrasekhar}
S. Chandrasekhar, The Mathematical Theory of Black Holes (Oxford University Press, New York, 1983).

\bibitem{Scully-metric-test}
M. O. Scully, M. S. Zubairy, and M. P. Haugan, Proposed optical test of metric gravitation theories,
Phys. Rev. A {\bf 24}, 2009 (1981).

\bibitem{Boltzmann-BH_1}
J. B. Hartle and S. W. Hawking,
Path-integral derivation of black-hole radiance,
Phys. Rev. D {\bf 13}, 2188 (1976);
G. W. Gibbons and M. J. Perry,
Black Holes in Thermal Equilibrium,
Phys. Rev. Lett. {\bf 36}, 985 (1976).

\bibitem{Boltzmann-BH_2}
T. Padmanabhan,
Thermodynamical Aspects of Gravity: New insights,
Rep. Prog. Phys. {\bf 73}, 046901 (2010);
and references therein.

\bibitem{UA-mirror-Svidzinsky}
This is similar to a mirror uniformly accelerated relative to an inertial atom; see Ref.~\cite{scully2019} and:
A. A. Svidzinsky, J. S. Ben-Benjamin, S. A. Fulling, and D. N. Page,
Excitation of an atom by a uniformly accelerated mirror through virtual transitions,
Phys. Rev. Lett. {\bf 121}, 071301 (2018).
 
\bibitem{fulling2018}
S. A. Fulling and J. H. Wilson, The equivalence principle at work in radiation from unaccelerated atoms and mirrors, Physica Scripta {\bf 94}, 014004 (2018).

\bibitem{Kerr-vacuum-states_1}
 A. C. Ottewill and E. Winstanley, 
 Renormalized stress tensor in Kerr space-time: General results,
 Phys. Rev. D {\bf 62},  084018 (2000);
M. Casals and A. C. Ottewill
 Canonical quantization of the electromagnetic field on the Kerr background,
Phys. Rev. D {\bf 71}, 124016 (2005).

\bibitem{Kerr-vacuum-states_2} 
 G. Menezes, 
 Spontaneous excitation of an atom in a Kerr spacetime,
 Phys. Rev. D {\bf 95}, 065015 (2017).
 
 \bibitem{Chrzanowski}
P. L. Chrzanowski and C. W. Misner, 
 Geodesic synchrotron radiation in the Kerr geometry by the method of asymptotically factorized Green's functions,
Phys. Rev. D {\bf 10}, 1701 (1974).

\bibitem{Unruh-Starobinsky-rad}
A. A. Starobinsky, 
Amplification of waves during reflection from a rotating ``black hole,''
Sov. Phys. JETP {\bf 37}, 28 (1973);
W. G. Unruh, 
Second quantization in the Kerr metric. 
Phys. Rev. D {\bf 10}, 3194 (1974).


\bibitem{Kerr-fermion-Q}
M. Casals, S. R. Dolan, B. C. Nolan, A. C. Ottewill, and E. Winstanley,
Quantization of fermions on Kerr space-time,
Phys. Rev. D {\bf 87}, 064027 (2013);
V. Ambru\c{s} and E. Winstanley, 
Rotating quantum states,
Phys. Lett. B {\bf 73}, 296 (2014).

\end{thebibliography}
\end{document}